\documentclass{ptephy_v1}

\usepackage[latin1]{inputenc}
\usepackage{graphics}
\usepackage{amsmath,amsthm,amssymb}
\usepackage{bm}
\usepackage{physics}
\usepackage{aas_macros}

\usepackage{amsfonts}
\usepackage{dcolumn}

\newcommand*{\beqa}{\begin{eqnarray}}
\newcommand*{\eeqa}{\end{eqnarray}}

\renewcommand{\k}{\kappa}
\newcommand*{\p}{\partial}

\newcommand{\n}{{\nu}}
\newcommand{\m}{{\mu}}
\newcommand{\vp}{{\varphi}}
\newcommand{\barg}{\overline{g}}
\newcommand{\barn}{\overline{\nabla}}
\newcommand{\barR}{\bar{R}}

\renewcommand{\P}{{\Phi}}

 \newcommand{\simg}{\gtrsim}
\newcommand{\siml}{\lesssim}
 
\begin{document}

\title{Spontaneous Scalarization in Scalar-Tensor Theories with 
Conformal Symmetry as an Attractor}

\author{Takeshi Chiba}

\affil{Department of Physics, College of Humanities and Sciences, Nihon University, \\
                Tokyo 156-8550, Japan}


\begin{abstract}

Motivated by constant-G theory,  
we introduce a one-parameter family of scalar-tensor theories as an 
extension of constant-G theory in which 
the conformal symmetry is a cosmological attractor. 
Since  the model has the coupling function of negative curvature, we expect 
spontaneous scalarization occurs and the parameter is constrained by 
pulsar-timing measurements.  
Modeling neutron stars with realistic equation of states, 
we study the structure of neutron stars  and calculate 
the effective scalar coupling with the neutron star in these theories. 
We find that within the parameter region where the observational constraints are satisfied,  
the effective scalar coupling almost coincides with that derived using  
the quadratic model with the same curvature. 
This indicates that the constraints obtained by 
the quadratic model will be used to limit the curvature of the coupling function 
universally in the future.   
\end{abstract}


\maketitle

\section{Introduction}

The equivalence principle played the principal role in constructing 
general relativity (GR) by Einstein.  
The weak equivalence principle (WEP) states that 
the motion of a (uncharged) test body is independent of its internal structure 
and composition. 
 WEP together with the local Lorentz invariance 
(independence of the results of nongravitational experiments from 
the velocity of the local Lorentz frame) 
and the local position invariance 
(the independence of the experimental results from the spacetime position) 
enables us to make the matter couple universally to gravity.  
Among various gravity theories, GR is the exception in that the equivalence principle is satisfied even for self-gravitating bodies: the strong equivalence principle (SEP).

In general scalar-tensor theories of gravity \cite{Bergmann:1968ve,Nordtvedt:1970uv,Wagoner:1970vr},  the equation of motion of a self-gravitating massive body  
(at the first post-Newtonian approximation) depends on the inertial mass $m_I$ 
and the (passive) gravitational mass $m_G$ of the body. The ratio is given by
\beqa
\frac{m_G}{m_I}=1+\eta\frac{\Omega}{m_I}
\eeqa
where $\Omega$ is the gravitational self-energy of the body \cite{will2018}. 
This implies that the motion of a massive body depends on its internal structure,  
being in violation of  the SEP. 
The coefficient in front of the gravitational self-energy is $\eta=4\beta-\gamma-3$, where $\beta$ and $\gamma$ are the parametrized post-Newtonian (PPN) parameters. For $\eta\neq 0$, the orbit of the  Moon around the Earth 
is elongated  toward the Sun (Nordtvedt effect \cite{Nordtvedt:1968qs}).\footnote{ 
{}From the lunar-laser-ranging experiment, $\eta$ is constrained as $\eta=(-0.2\pm1.1)\times 10^{-4}$ \cite{Hofmann:2018myc}.} 

In GR, $\beta=\gamma=1$ so that the equation of motion solely depends on 
the inertial mass and the velocity of its  
center of mass. However, even in scalar-tensor gravity, 
one may construct a theory with $\eta=0$ so that SEP holds. 
The ``constant-$G$'' theory by Barker \cite{1978ApJ...219....5B} is such an example. 
In fact, as shown by \cite{Damour:1992we}, it is the unique 
scalar-tensor theory which satisfies the SEP (at the first post-Newtonian approximation). 
Even more interestingly,  the theory is 
cosmologically attracted toward the conformally symmetry \cite{Gerard:2006ia} where the theory is scale-invariant. 
The conformal symmetry (and its spontaneous breaking) 
has been actively studied in constructing models of 
inflation \cite{Kallosh:2013hoa,Kallosh:2013daa} 
and in constructing geodesically complete cosmologies \cite{Bars:2011aa,Bars:2013yba}. 
Here the conformal symmetry will be restored in the future.  
Motivated by this remarkable property of constant-$G$ theory, we introduce 
a one-parameter family of scalar-tensor theories which exhibits 
the cosmological attraction toward  conformal symmetry.

Since the curvature of the coupling function of these theories is  negative,     
the scalar field experiences a tachyonic instability inside a compact 
star like a neutron star and the scalar field exhibits large deviation from 
its asymptotic value, a phenomenon so-called 
``spontaneous scalarization'' \cite{Damour:1993hw}, which is constrained by 
pulsar-timing observations.

In order to put constraints on the scalar-tensor theories from the binary pulsars, 
the dependence of the scalarization both on the equation of state (EOS) of a 
neutron star and on the coupling function (the function which determines 
the strength of the coupling to matter in the Einstein frame) must be taken into account. 
The dependence of the scalarization on the EOSs is 
studied in \cite{Harada:1997mr,Novak:1998rk,Damour:1998jk,Shibata:2013pra,Shao:2017gwu,Wex:2020ald}, and it is found that the threshold value of the scalarization is insensitive to EOS.  Moreover, Ref.\cite{Shao:2017gwu} find 
that there are several observational windows   
for the scalarization depending on the EOSs. 
However, the dependence on the coupling function 
has not been much studied and  usually the quadratic model \cite{Damour:1993hw}  is used. 
\cite{Anderson:2019eay} studied binary pulsar constraints on 
two scalar-tensor theories (the quadratic model \cite{Damour:1993hw} and MO model 
\cite{Mendes:2016fby}) 
and found that the constraints in the two theories are roughly the same. 

We extend these previous studies by using our coupling function 
in light of new and updated pulsar data. We find that  the  window at $\sim 1.7 M_{\odot}$ 
is almost closed even if the dependence of the EOSs is taken into account. 
The coupling function used in this study deviates from a quadratic function 
with the same curvature for a large scalar field. 
However, we find that the deviation is  small within  
the parameter region where the observational constraints are satisfied. 
Therefore, as far as the observational constraints on spontaneous scalarization are concerned, it is sufficient to employ the quadratic function 
for the coupling function. 

The paper is organized as follows. In Sec. \ref{sec2}, we review the properties of 
constant-G theory and introduce the conformal attractor model, a one-parameter family 
of scalar-tensor theories with the cosmological attraction toward the conformal symmetry. 
In Sec. \ref{sec3}, we study the structure of neutron stars in these theories 
and calculate 
the masses of neutron stars and the effective scalar coupling with the neutron star 
using three EOSs and compare them with the observational constraints. 
We also calculate the effective scalar coupling for the quadratic model. 
Sec. \ref{sec4} is devoted to summary.  We use the units of $c=1$.

\section{Cosmological Conformal Attractor}
\label{sec2}

\subsection{$\eta$ and Constant-$G$ Theory}

We consider scalar-tensor theories of gravity whose action in 
the Jordan frame is given by
\beqa
S=\int d^4x\sqrt{-g}\frac{1}{16\pi}\left(\Phi R-\frac{\omega(\P)}{\P}(\nabla\P)^2\right)+S_M(\psi,g_{\mu\nu})\,,
\label{action}
\eeqa
where $\P$ is the so-called Brans-Dicke scalar field, the inverse of which 
plays the role of  the effective gravitational ``constant'', 
$\omega(\P)$ is the Brans-Dicke function which determines the strength 
of the coupling  of the scalar field to gravity (and matter), 
$S_M$ is the matter action and $\psi$ denotes the matter field.  

We note that for $\omega=-3/2$, the gravity part of the action 
is locally conformal invariant 
under the following Weyl scaling\footnote{The action is still conformal invariant even if 
one includes a potential proportional to $\P^2$. However, the equation of the motion of the scalar field is not affected by such a potential 
 as discussed in \ref{sec21}. }:
\beqa
g_{\mu\nu}\rightarrow e^{2\sigma(x)}g_{\mu\nu},~~~~~\P\rightarrow e^{-2\sigma(x)}\P\,.
\eeqa

{}From the post-Newtonian expansion of the theory, 
the effective gravitational constant $G$ and 
the parametrized post-Newtonian (PPN) parameters 
$\gamma$ and $\beta$ of scalar-tensor theories are given by \cite{will2018}\footnote{We do not use the units of the present-day gravitational constant $G=1$ at this stage 
because we are interested in the cosmological 
evolution of $\P$. }
\beqa
G&=&\frac{1}{\P}\frac{2\omega(\P)+4}{2\omega(\P)+3}\bigg{|}_{\P_0}\label{effective-G}\\
\gamma&=&\frac{\omega+1}{\omega+2}\,,
\label{gamma}\\
\beta&=&1+\frac{\P\frac{d\omega}{d\P}}{4(2\omega+3)(\omega+2)^2}\label{beta}\,.
\eeqa
The equation of motion of a massive body is given in \cite{Damour:1992we,will2018}. It contains terms which depend on 
 the gravitational self-energy, the coefficient of which is $\eta=4\beta-\gamma-3$. For $\eta= 0$, the motion of massive bodies does not 
 depend on their internal structure and hence respects SEP. 
 
Remarkably, $\eta$ and $G$ are related by (see \cite{Damour:1992we} for a similar relation)
\beqa
\eta=(\gamma-1)\frac{d\ln G}{d\ln \P}.
\eeqa
Therefore, $\eta=0$, for which the theory respects the SEP, implies 
either $\gamma=1$ (GR) or $G={\rm const}$. The latter is known as 
``constant-$G$'' theory found by Barker \cite{1978ApJ...219....5B} in which 
$\omega(\P)$ is given by
\beqa
\omega(\P)=-\frac32+\frac{1}{2G\P-2}
\label{omega}
\eeqa
Note that $|\omega|\rightarrow \infty$ (GR) for $G\P\rightarrow 1$ and that 
$\omega\rightarrow -3/2$ (conformal symmetry) for $G|\P|\rightarrow \infty$. 
The constraint on $\gamma$ from the measurement of the time delay of 
the Cassini spacecraft is $\gamma-1=(2.1\pm 2.3)\times 10^{-5}$ \cite{Bertotti:2003rm}, 
which is satisfied for $G\P-1=(-1.1\pm 1.2)  \times 10^{-6}$ at present.  

\subsection{Cosmological Evolution}
\label{sec21}

In order to study the dynamics of $\P$, it is useful to perform 
the following change of variables and moving to 
the so-called Einstein frame \cite{Damour:1993id,Chiba:2013mha} in which $\P$ is 
decoupled from the gravity sector: 
\beqa
g_{\mu\nu}&=&\frac{1}{G_*\P}\barg_{\mu\nu}\equiv e^{2a(\vp)}\barg_{\mu\nu}
\label{conf-factor}\\
\frac{1}{2\omega+3}&=&
\frac{1}{4\pi G_*}\left(\frac{da(\vp)}{d\vp}\right)^2
\equiv\frac{1}{4\pi G_*}\alpha(\vp)^2
\label{scalar:vp}
\eeqa
where $G_*$ is the bare gravitational constant. 
The action (\ref{action}) can be rewritten  in terms of $\barg_{\m\n}$ whose kinetic term is of Einstein-Hilbert form and a canonically normalized scalar field $\vp$ as \footnote{
Note that our $\vp$ is related to the scalar field  in \cite{Damour:1992we,Damour:1993hw}  $\vp_{\rm DEF}$ via $\vp_{\rm DEF}=\k\vp/\sqrt{2}$. }
\beqa
S=\int d^4x\sqrt{-\barg}\left(\frac{\barR}{16\pi G_*}-\frac12 (\barn\vp)^2\right)
+S_M(\psi,e^{2a}\barg_{\mu\nu})
\label{action2}
\eeqa 
{}From Eq. (\ref{action2}), the equation of motion of $\barg_{\m\n}$ and $\vp$ are given by
\beqa
&&{\overline R}_{\m\n}-\frac12 \barg_{\m\n}{\overline R}=8\pi G_*
\left({\overline T}_{\m\n}+\p_{\m}\vp\p_{\n}\vp-\frac12\barg_{\m\n}
(\barn\vp)^2\right)
\label{st:einstein}\\
&&\overline\Box \vp=-\alpha(\vp){\overline T}
\label{scalar:einstein}
\eeqa
where ${\overline T}_{\m\n}=-(2/\sqrt{-\barg})\delta S_M/\delta \barg^{\m\n} $ is the energy-momentum tensor in the Einstein frame. From Eq. (\ref{scalar:einstein}), 
we find that $\vp$ moves according to the effective potential $-a(\vp){\overline T}$.\footnote{We note that 
the action Eq.(\ref{action}) is still conformal invariant even if 
one includes a potential proportional to $\P^2$. In the Einstein frame action Eq.(\ref{action2}) such a potential corresponds to a constant and does not affect 
the evolution of the scalar field. }
The effective gravitational constant $G$ and the PPN parameter $\gamma$ are written as
\beqa
G&=&G_*e^{2a(\vp)}\left(1+2\alpha(\vp)^2/\kappa^2\right)\bigg{|}_{\vp_0}\\
\gamma-1&=&-\frac{4\alpha(\vp)^2/\kappa^2}{2\alpha(\vp)^2/\kappa^2+1}\bigg{|}_{\vp_0}
\label{gamma-e}
\eeqa
where $\kappa=\sqrt{8\pi G_*}$ and $\vp_0$ denotes the value of $\vp$ at spatial infinity.

\begin{figure}[htp]
	\centering
	\includegraphics[width=0.55\textwidth]{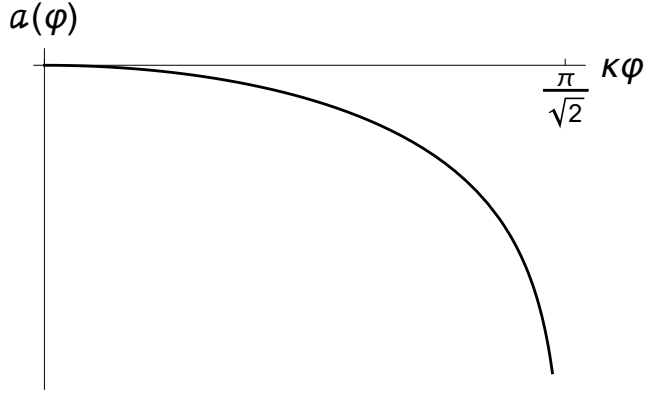}
	\caption{ \label{fig1}
	$a(\vp)$ in Eq. (\ref{conf-factor2}). $\k\vp=0$ corresponds to GR and 
	$\k\vp=\pi/\sqrt{2}$ corresponds to the conformal symmetry.  } 
\end{figure}

For constant-$G$ theory with $\omega(\P)$ in Eq. (\ref{omega}),  
{}from Eq. (\ref{conf-factor}) and Eq. (\ref{scalar:vp}),  
$\vp$ and $a(\vp)$ are given by 
\beqa
\k\vp&=&\sqrt{2}\tan^{-1}\sqrt{G\P-1}\\
a(\vp)&=&\ln\left(\cos\left(\k\vp/\sqrt{2}\right)\right)
\label{conf-factor2}
\eeqa
where $G\P \geq 1$ is assumed ($\cosh$ for $G\P<1$) and $G_*$ coincides with $G$. 
We have 
fixed the integration constant so that $G\P=1$ corresponds to $\k\vp=0$.  
The conformal symmetry 
($G\P\rightarrow \infty$) now 
corresponds to $\k\vp\rightarrow \pi/\sqrt{2}$.  In Fig. \ref{fig1}, $a(\vp)$ is shown. 
The constraint by the Cassini experiment is satisfied 
for $\k\vp=(-2.6\pm 4.0)\times 10^{-3}$ at present. 
{}From Eq. (\ref{scalar:einstein}), one may see that $\vp$ 
moves toward $\k\vp\rightarrow \pi/\sqrt{2}$ according to the effective potential 
$-a(\vp){\overline T}$ (see Fig. \ref{fig1}) during 
the matter-dominated epoch and during the dark energy-dominated epoch: 
The theory is thus cosmologically attracted toward the conformal symmetry \cite{Gerard:2006ia}.

\subsection{Conformal Attractor Model}

Although the scalar-tensor theory with $\eta=0$ is unique, 
we may consider possible generalization of the models 
which exhibit cosmological attraction toward the conformal symmetry. 

For example, we can consider the following generalization of the coupling function  $a(\vp)$ 
in Eq. (\ref{conf-factor2}):
\beqa
a(\vp)=\ln \left(\cos\left(\sqrt{p}~\k\vp\right)\right)
\label{factor-ext}
\eeqa
where  $p$ is a non-negative parameter and 
$p=1/2$ corresponds to constant-G theory. 
{}From Eq. (\ref{conf-factor}) and Eq. (\ref{scalar:vp}), the corresponding $\P$ and $\omega(\P)$ are given by
\beqa
G_*\P&=&\sec^{2}(\sqrt{p}~\k\vp)\\
\omega(\P)&=&-\frac32 +\frac{1}{4p}\frac{1}{G_*\P-1}\,.
\eeqa
 The conformal symmetry $\omega=-3/2$  now corresponds to $\k\vp=\pi/(2\sqrt{p})$ 
  and the shape of $a(\vp)$ is similar to Eq. (\ref{conf-factor2}) ($p=1/2$) 
 and we expect similar cosmological attraction toward the conformal symmetry. \footnote{We note that a theory with a quadratic function $a(\vp)=-\frac12 p (\k\vp)^2$ with $p>0$ 
 \cite{Damour:1993hw}  is 
 also cosmologically attracted toward the conformal symmetry.  } 

On the other hand, the effective gravitational constant  $G$  defined
 by Eq. (\ref{effective-G}) is given by 
\beqa
G=2pG_* +\frac{1-2p}{\P}, 
\eeqa
and the gravitational constant is no longer constant. 
PPN parameters $\gamma$ and $\beta$ and $\eta=4\beta-\gamma-3$ 
are given from Eq.(\ref{gamma}) 
and Eq. (\ref{beta}) by 
\beqa
\gamma-1&=&-\frac{4p(G_*\P-1)}{2pG_*\P+1-2p}\\
\beta-1&=&-\frac{2p^2G_*\P(G_*\P-1)}{(2pG_*\P+1-2p)^2}\\
\eta&=&-\frac{4p(2p-1)(G_*\P-1)}{(2pG_*\P+1-2p)^2}.
\eeqa
In terms of $\gamma$, $\eta$ can be rewritten in a suggestive form:
\beqa
\eta=(2p-1)\left(\frac{\gamma+1}{2}\right)(\gamma-1).
\eeqa
Therefore, although $\eta$ is no longer vanishing  as long as $p\neq 1/2$, 
it is suppressed by $\gamma-1$. 
From the bound by Cassini $\gamma-1=(2.1\pm 2.3)\times 10^{-5}$ \cite{Bertotti:2003rm},  the constraint on $\eta$ from 
the lunar-laser-ranging experiment, $\eta=(-0.2\pm1.1)\times 10^{-4}$ \cite{Hofmann:2018myc}, is satisfied for $p<2.9$.

\section{Spontaneous Scalarization in Conformal Attractor Model}
\label{sec3}

The curvature  of the coupling function 
Eq. (\ref{factor-ext}) at $\vp=0$ is negative.    
For  $a(\vp)$ with negative curvature, 
the scalar field experiences a tachyonic instability inside a compact 
star like a neutron star and exhibits a large deviation from 
its asymptotic value, a phenomenon so-called 
``spontaneous scalarization'' \cite{Damour:1993hw}.  From 
the observation of the pulsar-timings of several neutron star-white dwarf 
binaries, such a phenomenon has been constrained \cite{Antoniadis:2013pzd,Shao:2017gwu}.  

Assuming a quadratic form of $a(\vp)=-\frac12 p(\k\vp)^2$, $p$ is constrained as 
$p\siml 2.2.$\footnote{Note that 
since $\vp_{\rm DEF}$ in \cite{Damour:1993hw}  corresponds to  
$\vp_{\rm DEF}=\k\vp/\sqrt{2}$, $\beta_{\rm DEF}$ in $a(\vp_{\rm DEF})=\frac12\beta_{\rm DEF}\vp_{\rm DEF}^2$ corresponds to $\beta_{\rm DEF}=-2p$. }
Although the coupling function $a(\vp)$ in Eq. (\ref{factor-ext}) is well approximated 
as $a(\vp)=-\frac12 p(\k\vp)^2$ for  $\k\vp\ll 1$,  
the scalar field 
acquires a large value when the theory exhibits  spontaneous scalarization and 
higher order terms in the Taylor expansion of $a(\vp)$ may not be negligible and 
 it is not clear whether the limit on $p$ from the quadratic model may apply here. 

Hence,  in this section, 
we study  spontaneous scalarization for the conformal attractor model 
Eq. (\ref{factor-ext}) and compare it with the quadratic model. 

\subsection{TOV equation}

We study the spherically symmetric static solutions generated by perfect 
fluid neutron stars in scalar-tensor theories.  
The metric in the Einstein frame is assumed to be of the form
\beqa
\barg_{\m\n}dx^{\mu}dx^{\nu}=-e^{\nu(r)}dt^2+\frac{dr^2}{1-2\mu(r)/r}+r^2d\Omega^2\,.
\eeqa
The perfect fluid energy-momentum in the Jordan frame is
\beqa
T^{\mu\nu}=(\epsilon+p)u^{\mu}u^{\nu}+pg^{\mu\nu}
\eeqa
and is related to the energy-momentum tensor in the Einstein frame by 
${{\bar T}^{\mu}}_{~~\nu}=e^{4a}{T^{\mu}}_{\nu}$. The equation of motion 
Eq. (\ref{st:einstein}) and Eq. (\ref{scalar:einstein}) become
\beqa
\mu'&=&4\pi G_*r^2e^{4a(\vp)}\epsilon+2\pi G_*r(r-2\mu)\vp'^2\label{metric1}\\
\nu'&=&\frac{2\mu}{r(r-2\mu)}+8\pi G_*\frac{r^2e^{4a(\vp)}}{r-2\mu}p+4\pi G_*r\vp'^2\label{metric2}\\
\vp''&=&-\frac{2(r-\mu)}{r(r-2\mu)}\vp'+4\pi G_*\frac{r^2e^{4a(\vp)}}{r-2\mu}(\epsilon-p)\vp'
+\frac{re^{4a(\vp)}}{r-2\mu}(\epsilon-3p)\alpha(\vp)\label{scalar-evolv}\\
p'&=&-(\epsilon+p)\left(\frac{\mu}{r(r-2\mu)}+4\pi G_*\frac{r^2e^{4a(\vp)}}{r-2\mu}p
+2\pi G_*r\vp'^2+\alpha(\vp)\vp'\right)
\eeqa
where the prime denotes a derivative with respect to $r$.  

With the coupling function $a(\vp)$ Eq. (\ref{factor-ext}), given  
the initial conditions at $r=0$, we numerically integrate these equation 
outward using a 4th-order Runge-Kutta method. 
The pressure goes to zero at the surface of star, and beyond that only 
the metric and scalar equations  with vanishing $p$ and $\epsilon$ are necessary. 
Specifically, we set $p$ and $\epsilon$ to zero if $p$ is 
less than $10^{-7}m_Bn_0$ (with $m_B=1.66\times 10^{-24}{\rm g}$ and $n_0= 0.1 {\rm fm}^{-3}$)  well below  neutron drip 
and integrate Eq. (\ref{metric1}) $\sim$ Eq. (\ref{scalar-evolv}) further outward. 
The regularity at $r=0$ requires $\mu(0)=\nu(0)=\vp'(0)=0$.  
We can freely specify $\vp(0)$ and $p(0)$. But in order to satisfy the 
bound  by  the Cassini satellite experiment $|\gamma-1|<2.3\times 10^{-5}$, 
from Eq. (\ref{gamma-e}) the asymptotic value $\vp_0$ should satisfy 
$\alpha(\vp_0)/\k<2.4\times 10^{-3}$. 
Considering the limit on $\gamma$ and  
in order to put conservative constraints on the theory, 
we require $\alpha(\vp_0)/\k=1.0\times 10^{-4}$ at large $r$. Hence, 
for a given $p(0)$, a particular value of $\vp(0)$ can satisfy this condition. 
We employ the shooting method to find $\vp(0)$.

\subsection{Equation of State}

As an equation of state (EOS), 
we adopt piecewise-polytropic parametrizations for the nuclear EOS by 
Read {\it et al.} \cite{Read:2008iy} for APR4 \cite{Akmal:1998cf} and 
H4 \cite{Glendenning:1991es,Lackey:2005tk} and MS1 \cite{Mueller:1996pm} EOSs. 
APR4  (variational method) and MS1 (relativistic mean-field theory) are EOSs for 
nuclear matter composed of neutrons, protons, electrons, and muons, whereas H4 
(relativistic mean-field theory) includes the effect of hyperons in addition. 

A piecewise polytropic EOS consists of several polytropic EOSs
\beqa
p(\rho)=K_i\rho^{\Gamma_i},~~~~~~\rho_{i-1}\leq \rho\leq \rho_i,
\eeqa
where $\rho$ is the rest-mass density. From the continuity of the pressure at $\rho_i$ determines $K_{i+1}$ at the next interval as 
$K_{i+1}={p(\rho_i)}/{\rho_i^{\Gamma_{i+1}}}$. 
  The energy density $\epsilon$ is determined by 
the first law of thermodynamics, $d(\epsilon/\rho)=-pd(1/\rho)$ as
\beqa
\epsilon=(1+a_i)\rho+\frac{K_i}{\Gamma_i-1}\rho^{\Gamma_i},~~~~~\rho_{i-1}\leq \rho\leq \rho_i,
\eeqa
where $a_i$ is an integration constant given by
\beqa
a_i=\frac{\epsilon(\rho_{i-1})}{\rho_{i-1}}-1-\frac{K_{i-1}}{\Gamma_i-1}\rho_{i-1}^{\Gamma_i-1}
\eeqa
{}from the continuity of the energy density at $\rho_{i-1}$.


In the four-parameter model  of \cite{Read:2008iy}, the EOS at low densities 
(crust EOS) is fixed to the EOS of Douchin and Haensel \cite{Douchin:2001sv} and 
is matched  to a polytrope with adiabatic exponent $\Gamma_1$. At a fixed 
rest-mass density\footnote{In terms of the nuclear saturation density 
$\rho_{\rm nuc}\simeq 2.7\times 10^{14}{\rm g/cm^3}$, 
$\rho_1\simeq 1.9 \rho_{\rm nuc}$. } 
$\rho_1=10^{14.7} {\rm g/cm^3}$ and pressure $p_1=p(\rho_1)$, the EOS is 
joined continuously to a second polytrope  with $\Gamma_2$. Finally, at 
$\rho_2=10^{15}{\rm g/cm^3}$, the EOS is joined to a 
third polytrope with $\Gamma_3$. 
Further details are described in \cite{Read:2008iy}.

\begin{figure}[htp]
	\centering
	\includegraphics[width=0.75\textwidth]{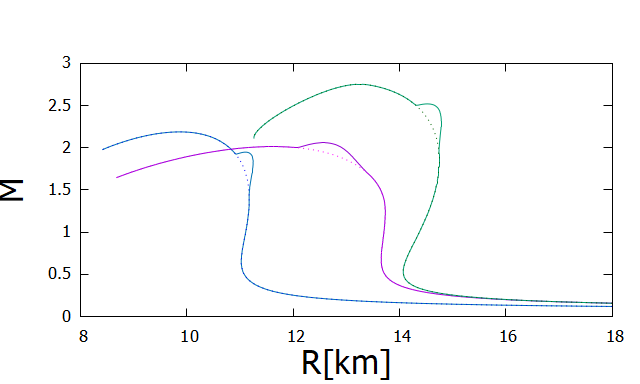}
	\caption{ \label{fig20}
	Mass-radius relation for APR (blue, left), H4(magenta, middle), MS1(green, right) EOSs. The 
	solid curves are for  the conformal attractor model Eq. (\ref{factor-ext})	 
	with $p=2.3$ and the dotted  curves are for GR. 
	  } 
\end{figure}

In Fig. \ref{fig20}, we show the mass-radius relation for three EOSs for the conformal attractor model Eq. (\ref{factor-ext})  
with $p=2.3$ (solid) and GR (dotted).  The physical radius of a neutron star $R$ is 
defined in terms of the surface of the star $r_*$ by $R=e^{a(\vp(r_*))}r_*$. 
Among three EOSs, MS1 is the stiffest 
(large $p_1$) EOS and the radius of a $1.4 M_{\odot}$ neutron star is 
large ($\sim 14.6 {\rm km}$ in GR), while  APR4 is a soft EOS and 
the radius of a $1.4 M_{\odot}$ neutron star is small ($\sim 11.2 {\rm km}$). 
In fact, a strong correlation between the pressure at around the  nuclear saturation 
density $\rho_{\rm nuc}$ and the radius of $1.4 M_{\odot}$ neutron stars 
has been found \cite{Lattimer:2000nx}. 
H4 is in between the two EOSs.  On the other hand, the maximum 
mass of a neutron star (in GR) is the largest for MS1 ($\sim 2.75 M_{\odot}$), while 
the smallest   for H4  ($\sim 2.01 M_{\odot}$) but is still compatible 
with the most massive pulsar ($2.08\pm 0.07 M_{\odot}$) measured by \cite{NANOGrav:2019jur,Fonseca:2021wxt}.  

\begin{figure}[htp]
	\centering
	\includegraphics[width=0.49\textwidth]{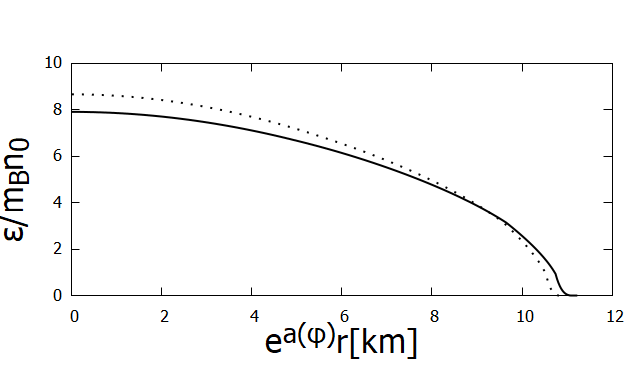}
	\includegraphics[width=0.49\textwidth]{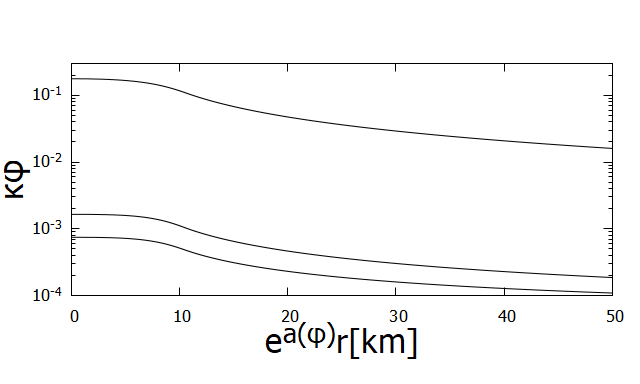}
	\caption{ \label{fig21}
	Left: The energy density $\epsilon$ (in unit of $m_Bn_0=1.66\times 10^{14}{\rm g/cm^3}$) as a function of 
	the physical radius $e^{a(\vp)}r$ for APR4 EOS. A solid curve is for the conformal attractor model with $p=2.3$ and  a dotted curve is for GR. Right: 
	the profile of the scalar field $\phi$. $p=2.3,2.2,2.1$ from top to bottom. 
	All cases have a fixed gravitational mass of $1.9 M_{\odot}$. 
	  } 
\end{figure}

We also show the radial profile of the energy density $\epsilon$ 
in unit of $m_Bn_0=1.66\times 10^{14}{\rm g/cm^3}$ for APR4 EOS 
in the left of Fig. \ref{fig21} and 
the radial profile of the scalar field $\vp$ in the right.

For each EOS, we compute the total gravitational (ADM) mass $M_A$ of the neutron star which is easily read off from the asymptotic behavior of $\mu(r)$ at infinity as  
$G_*M_A=\lim_{r\rightarrow \infty}\mu(r)$.  Also, from the asymptotic behavior of $\vp$, 
$\k\vp\rightarrow \k\vp_0+G_*M_S/r$, we compute the scalar charge $M_S$ 
of the neutron star. The effective scalar coupling $\alpha_A$ introduced in
 \cite{Damour:1992we,Damour:1993hw} is written in terms of $M_A$ and $M_S$ as 
 (note again that our $\vp$ is related to the scalar field $\vp_{\rm DEF}$ in 
\cite{Damour:1992we,Damour:1993hw}  via $\vp_{\rm DEF}=\k\vp/\sqrt{2}$) 
\beqa
\alpha_A=\frac{\sqrt{2}}{\kappa}\frac{\p \ln M_A}{\p\vp}=-\frac{M_S}{\sqrt{2}M_A}.
\eeqa


\subsection{Pulsar Constraints}

\begin{table}[htp]
\begin{center}
\begin{tabular}{lllllc}
\hline
\hline
Pulsar & Orbital period $P_b$(d) & $\dot P_b({\rm 10^{-12}s~s^{-1}})$ & Companion mass & Pulsar mass &$\alpha_A$\\
\hline
J0348+0432 &0.102424062722(7) & -0.273(45) &0.172(3) &2.01(4) & $<5.0 \times 10^{-3}$\\
J1738+0333 &0.3547907398724(13) &-0.0170(31) &0.181($^{+0.008}_{-0.007}$) &1.46($^{+0.06}_{-0.05}$) &$<2.7 \times 10^{-3}$ \\
J1012+5307&0.60467271355(3)&-0.061(4)&0.165(15)&1.72(16)&$<4.7\times 10^{-3}$ \\
J1713+0747 &67.8251299228(5)&-0.34(15)&0.290(11)&1.33(10)&$<3.4\times 10^{-3}$\\
J2222-0137&2.44576437(2)&-0.2509(76)&1.319(4)&1.831(10)&$<5.0\times 10^{-3}$\\
J1909-3744 &1.533449474305(5)&-0.51087(13)&0.209(1)&1.492(14)&$<4.0\times 10^{-3}$\\
\hline
\hline
\end{tabular}
\caption{
Parameters of NS-WD binaries, PSRs J0348+0432\cite{Antoniadis:2013pzd}, J1738+0333\cite{2012MNRAS.423.3328F}, J1012+5307\cite{2009MNRAS.400..805L,
2020MNRAS.494.4031M,2020ApJ...896...85D}, J1713+0747\cite{2019MNRAS.482.3249Z}, 
and J2222-0137\cite{2017ApJ...844..128C,Guo:2021bqa}, J1909-3744\cite{2020MNRAS.499.2276L} 
and the 2$\sigma$ limits on $\alpha_A$. 
\label{table2}}
\end{center}
\end{table}

We consider the following 6 NS-WD binaries: 
PSRs  J0348+0432\cite{Antoniadis:2013pzd}, J1738+0333\cite{2012MNRAS.423.3328F}, J1012+5307\cite{2009MNRAS.400..805L,2020MNRAS.494.4031M,2020ApJ...896...85D}, J1713+0747\cite{2019MNRAS.482.3249Z}, 
and J2222-0137\cite{2017ApJ...844..128C,Guo:2021bqa}, J1909-3744\cite{2020MNRAS.499.2276L}.

\begin{figure}[htp]
	\centering
	\includegraphics[width=0.75\textwidth]{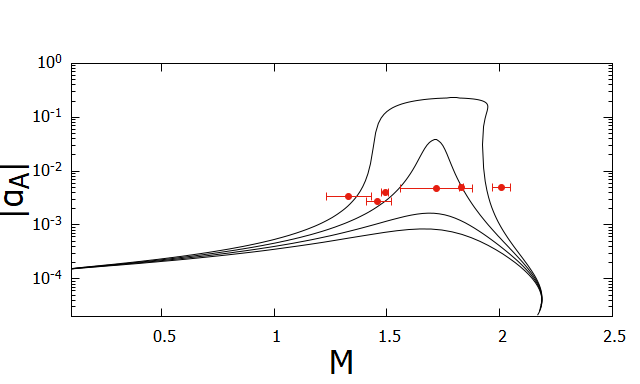}
	\includegraphics[width=0.75\textwidth]{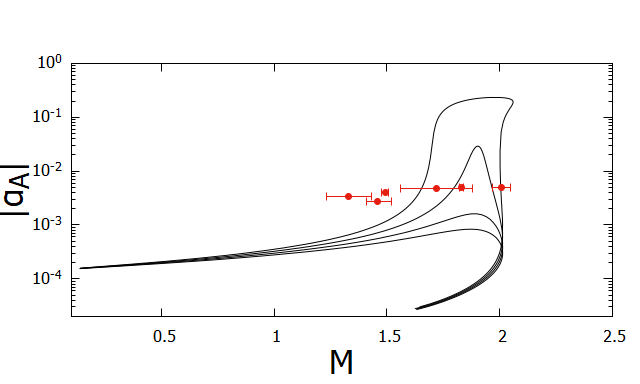}
	\includegraphics[width=0.75\textwidth]{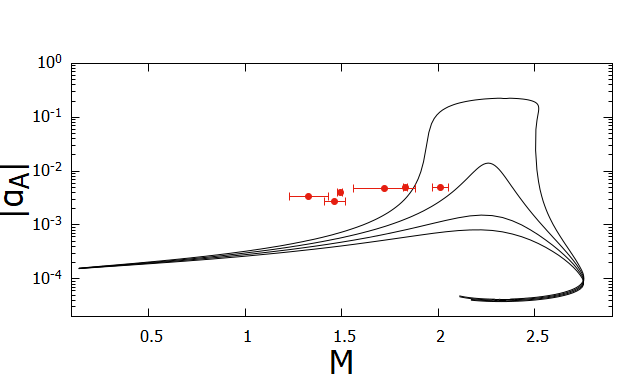}
	\caption{ \label{fig2}
	Effective scalar coupling $\alpha_A$ as a function of the gravitational mass of the neutron star for APR4(top), H4(middle) and MS1(bottom). The curves are for $p=2.0,2.1,2.2,2.3$ from bottom to top. 
	Red points are the 2$\sigma$ limits on $\alpha_A$. 
	  } 
\end{figure}

For PSR J1012+5307, the limit on the scalar coupling $\alpha_A$ from the absence of 
the dipole radiation is recently improved in 
\cite{2020ApJ...896...85D} by  
the precise measurement of the distance by VLBI. Moreover, the mass of the neutron star 
is determined recently due to an estimate of the mass of 
the white dwarf companion using binary evolution models \cite{2020MNRAS.494.4031M}. 
For PSR  J2222-0137, the limit on the scalar coupling  is recently improved in \cite{Guo:2021bqa} 
compared with  \cite{2017ApJ...844..128C} 
by the improved analysis of  VLBI data together with the extended timing data. 
For PSR 1713+0747, PSR J2222-0137 and PSR J1909-3744, the pulsar mass is determined from the Shapiro time-delays, while for others the pulsar mass is determined from the combination of the white dwarf mass determined from the optical spectrum and the mass ratio determined from the orbital velocity. 

The measurements of the orbital decay of the binary systems are consistent with 
the orbital decay due to the emission of gravitational waves predicted by GR \cite{Peters:1964zz}: 
\beqa
\dot P_b^{GR}=-\frac{192\pi}{5}\frac{\left(1+{73}e^2/24+{37}e^4/96\right)}{(1-e^2)^{7/2}}\left(\frac{2\pi}{P_b}\frac{G_*M_{c}}{c^3}\right)^{5/3}\frac{q}{(q+1)^{1/3}},
\eeqa 
where $e$ is the orbital eccentricity, $P_b$ is the orbital period, $M_c$ is the companion (white dwarf) mass and $q=M_A/M_c$ 
is the ratio of the pulsar mass to the companion mass. 
In scalar-tensor theory, scalar waves are also emitted and contribute the orbital decay. For the binary systems, the dominant contribution comes from dipolar waves \cite{Damour:1992we}:
\beqa
\dot P_b^D=-2\pi \frac{\left(1+e^2/2\right)}{(1-e^2)^{5/2}}\left(\frac{2\pi}{P_b}\frac{G_*M_c}{c^3}\right)\frac{q}{q+1}(\alpha_A-\alpha_c)^2,
\eeqa
where $\alpha_c$ are the effective scalar coupling to white dwarf. Since 
the self-gravity of white dwarf is small, $\alpha_c$ is no different 
from its asymptotic value: $\alpha_c\simeq \sqrt{2}\alpha(\vp_0)/\k\ll 1$. 
The measurements of $\dot P_b$ constrain the dipole contribution, hence 
$\alpha_A-\alpha_c\simeq \alpha_A$.  
The parameters of six NS-WD binaries and the $2 \sigma$ limits on $\alpha_A$ are 
shown in Table \ref{table2}.

In Fig.  \ref{fig2}, $\alpha_A$ as a function of the gravitational mass of the neutron stars for three EOSs together with the 2$\sigma$ limits on $\alpha_A$ from the pulsar-timings 
are shown. 
We find that spontaneous scalarization occurs if $p\simg 2.2$ and 
$p$ is constrained as  $p< 2.3$  irrespective of EOSs.  \footnote{The weak 
dependence of the onset of the scalarization on the EOS was found in \cite{Harada:1997mr,Novak:1998rk}. }
However, more detailed constraints on $p$ depend on the EOS: 
$p$ is constrained to be 
$p<2.2$ for APR4 and H4, while $p=2.2$ is allowed for MS1. 
We may place a conservative limit on $p$ to be $p<2.3$. 
Although there exist small windows for the scalarization 
at $\sim 1.9 M_{\odot}$ for H4 and at $\sim 2.3 M_{\odot}$ for MS1 \cite{Shao:2017gwu}, 
a ``window at $\sim 1.7 M_{\odot}$'' \cite{Shao:2017gwu} is now closed 
with the inclusion of PSR J1012+5307 and PSR J2222-0137.


\subsection{Choice of Coupling Function}

\begin{figure}[htp]
	\centering
	\includegraphics[width=0.5\textwidth]{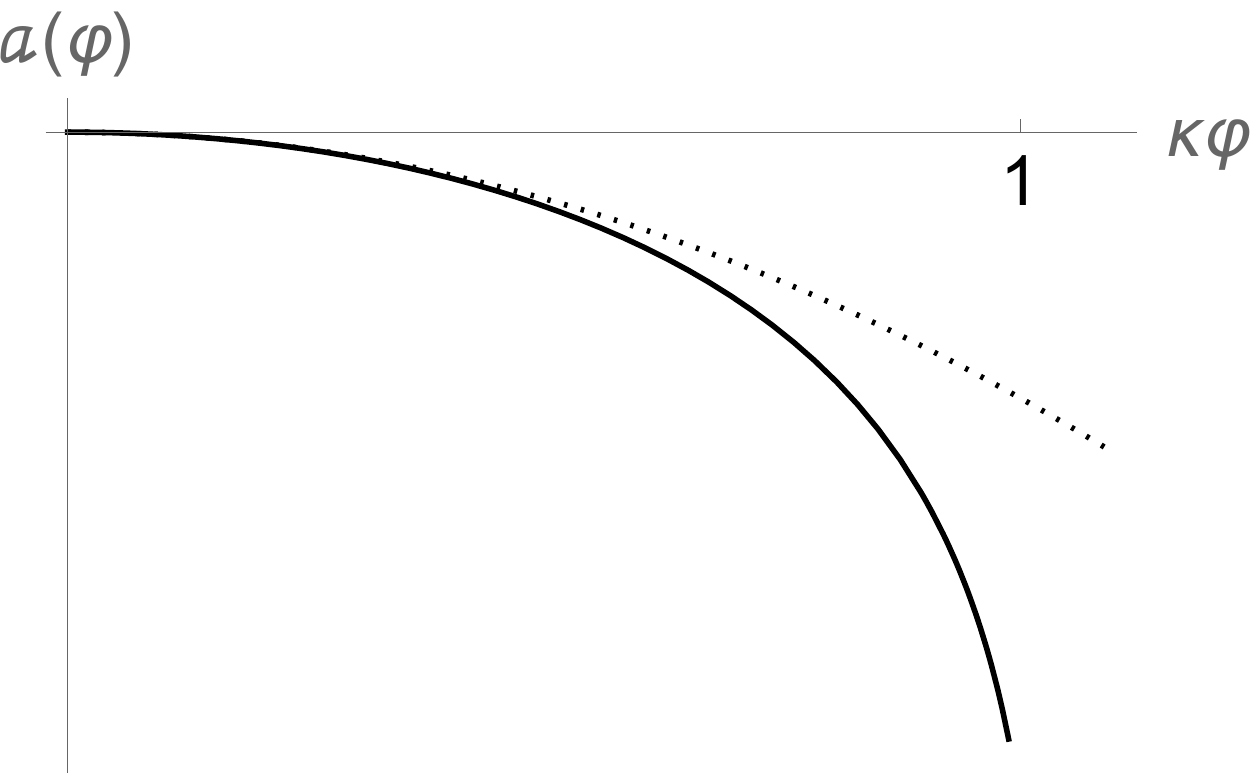}
	\caption{ \label{fig3}
	$a(\vp)$  of the conformal attractor model Eq. (\ref{factor-ext}) (solid line) and 
	of the 
	quadratic model $-\frac12p(\k\vp)^2$ (dotted line) 
	for $p=2.3$.  
	  } 
\end{figure}

So far, the effective scalar coupling is computed for 
the coupling function Eq. (\ref{factor-ext}). 
However, most frequently studied gravity theory is the so-called DEF 
theory \cite{Damour:1993hw} based on 
the quadratic coupling function : 
\beqa
a(\vp)=\frac12 \beta (\k\vp)^2.
\eeqa
Eq. (\ref{factor-ext}) can be expanded as $a(\vp)=-\frac12 p(\k\vp)^2+{ O}((k\vp)^4)$ for $\k\vp\ll 1$,
hence $p$ corresponds to $-\beta$ in the quadratic model.  
Note that 
since $\vp_{\rm DEF}$ in \cite{Damour:1993hw}  corresponds to  
$\vp_{\rm DEF}=\k\vp/\sqrt{2}$, $\beta_{\rm DEF}$ in $a(\vp_{\rm DEF})=\frac12\beta_{\rm DEF}\vp_{\rm DEF}^2$ corresponds to $\beta_{\rm DEF}=2\beta$. 

Although the coupling function $a(\vp)$ in Eq. (\ref{factor-ext}) is well approximated 
as $a(\vp)=-\frac12 p(\k\vp)^2$ as long as $\k\vp\ll 1$ (see Fig. \ref{fig3}), 
the scalar field may 
acquire a large value when the theory exhibits  spontaneous scalarization and 
the scalar field can experience a wider portion of the coupling 
function.

\begin{figure}[htp]
	\centering
	\includegraphics[width=0.75\textwidth]{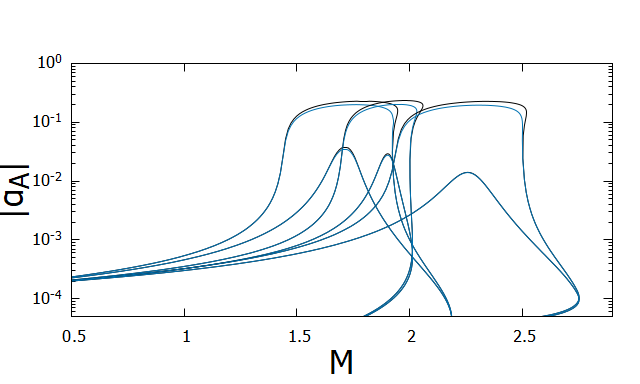}
	\caption{ \label{fig4}
	$\alpha_A$ for $a(\vp)$ in Eq. (\ref{factor-ext}) (black) and for $ a(\vp)=-\frac12 p (\k\vp)^2$ (blue) with $p=2.2,2.3$ (from bottom to top) for APR4, H4,MS1 (from left to right). 
	Two curves for $p=2.2$ overlap each other and are almost indistinguishable.  
	  } 
\end{figure}

\begin{figure}[htp]
	\centering
	\includegraphics[width=0.65\textwidth]{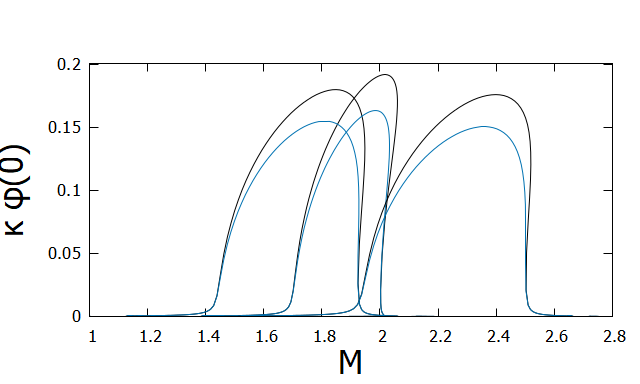}
	\caption{ \label{fig5}
	$\vp(0)$ for $a(\vp)$ in Eq. (\ref{factor-ext}) (black) and for $ a(\vp)=-\frac12 p (\k\vp)^2$ (blue) with $p=2.3$  for APR4, H4,MS1 (from left to right).
	  } 
\end{figure}

However, as shown in Fig. \ref{fig4}, even for $p=2.3$, 
the difference of $\alpha_A$ between 
the two coupling functions is very small.  For $p=2.2$, two $\alpha_A$s almost coincide. 
In fact, as shown in Fig. \ref{fig5}, 
the maximum value of the scalar field ($\vp(0)$) is at most $\k\vp(0)\siml 0.2$ even for $p=2.3$ and higher order terms in the Taylor expansion of $a(\vp)$ is negligible.  
Similar results are obtained by \cite{Anderson:2019eay} from the comparison of 
the quadratic model and  MO model \cite{Mendes:2016fby}.

This behavior can be understood from the stability analysis of neutron stars in scalar-tensor theories \cite{Harada:1997mr,Harada:1998ge}.  Spontaneous scalarization is  
triggered by a tachyonic instability of the scalar field of a general relativistic star. 
{}From the linear stability analysis of neutron stars,  the threshold value 
of the curvature of the coupling function is found to be insensitive 
to EOSs \cite{Harada:1997mr}. In the linear analysis, 
only the quadratic term in $a(\vp)$  is relevant 
and possible higher order terms are not important in the analysis.  
Moreover, using the catastrophe theory, it is shown 
that the onset of spontaneous scalarization in the quadratic coupling function corresponds to cusp catastrophe \cite{Harada:1998ge} which is structurally 
stable, i.e. the theory is stable against adding higher order terms in the coupling function. 
Therefore, we expect that the structure of the theory near the onset of 
the scalarization is described by the quadratic coupling function universally.

Since pulsar data already disfavor $p\geq 2.3$,  we thus conclude that 
as far as the observational constraints on spontaneous scalarization 
are concerned, it is sufficient to employ the quadratic function 
for the coupling function.  

\subsection{Comment on the Cosmological Evolution}

Finally we comment on the cosmological evolution of $\vp$ which defines its asymptotic value $\vp_0$. 

As we have seen in \ref{sec21}, the  conformal attractor model 
is cosmologically attracted toward the conformal symmetry $\omega\rightarrow -3/2$, 
in vast disagreement with the solar system experiments. Therefore, similar to the quadratic function with negative curvature, a severe fine-tuning of the initial conditions  
is required to satisfy the solar system constraints today\cite{Sampson:2014qqa}. 

One obvious and the simplest possibility to avoid this problem is to introduce 
a mass term for $\vp$ (in the Einstein frame)\cite{Chen:2015zmx,Ramazanoglu:2016kul,dePireySaintAlby:2017lwc}.  Massive scalar-tensor theories with mass $m_{\vp}$ 
of  $10^{-28}{\rm eV} \siml m_{\vp}\siml 10^{-10}{\rm eV}$ may both exhibit spontaneous scalarization (upper bound) and cosmological attraction toward GR 
in the matter-dominated era (lower bound). According to \cite{Ramazanoglu:2016kul}, 
the effective scalar coupling of a neutron star is not different from 
that of a massless theory 
if $m_{\vp}\siml 10^{-13}{\rm eV}$ but is much 
suppressed for $m_{\vp}\simg 10^{-12}{\rm eV}$.\footnote{
$m_{\vp}\siml 10^{-17}{\rm eV}$ may be required for the success of the big-bang nucleosynthesis \cite{dePireySaintAlby:2017lwc}.  If this is the case,  the structure of neutron stars would be almost the same as that of a massless theory.  } 


\section{Summary}
\label{sec4}

Motivated by constant-G theory which respects the SEP and 
is cosmologically attracted toward the conformal symmetry,  
we introduce a one-parameter family of scalar-tensor theories which exhibit  
cosmological attraction toward the conformal symmetry.  
From the constraint on the violation of SEP by the lunar-laser-ranging experiment, 
the parameter $p$ is constrained to $p<2.9$.

We have studied the structure of neutron stars in these theories for three realistic 
EOSs (APR4,H4,MS1). 
{}From the constraints on the effective scalar coupling from six neutron star-white 
dwarf binaries, the parameter $p$ is constrained to be $p<2.2$ for APR4 and H4, while 
$p<2.3$ for MS1. With new pulsar data a window for the scalarization at $\sim 1.7 M_{\odot}$  
is closed, but small windows are still open  at $\sim 1.9 M_{\odot}$ for H4 
and at $\sim 2.3 M_{\odot}$ for MS1. 

We have also compared in detail our coupling function and the quadratic model and 
have found that the difference of the effective scalar coupling 
between the two theories is very small for $p\leq 2.3$. 
Combining these results with the structural stability of the quadratic model 
at the onset of scalarization \cite{Harada:1998ge}, we conclude that  
as far as the observational constraints on spontaneous scalarization are concerned, 
the dependence on the coupling function is small and hence we can safely employ the quadratic function as the coupling function and the constraints obtained by 
the quadratic model will be used to put constraints on 
the curvature of the coupling function 
universally in the future.

\section*{Acknowledgments}
We would like to thank Masahide Yamaguchi for useful comments at 
the earlier stage of this work. 
This work is supported in part by Nihon University.

\bibliographystyle{JHEP}
\bibliography{references}

\end{document}